\documentclass[twocolumn,showpacs,groupedaddress,superscriptaddress,amsmath,amssymb]{revtex4}

\usepackage{graphicx}
\usepackage{dcolumn}
\usepackage{bm}

\begin{document}
\newcommand{\N}{\emph{N}}
\newcommand{\A}{N${}_2$}


\title{Cancellation of light-shifts in an \emph{N}-resonance clock}

\author{I. Novikova}
\affiliation{Harvard-Smithsonian Center
for Astrophysics, Cambridge, Massachusetts, 02138, USA}

\author{A.V. Taichenachev}
\author{V.I. Yudin}
\affiliation{Institute of Laser Physics SB RAS and Novosibirsk State
University, Novosibirsk, 630090, Russia}

\author{D.F. Phillips}
\affiliation{Harvard-Smithsonian Center for Astrophysics, Cambridge,
Massachusetts, 02138, USA}

\author{A.S. Zibrov}
\affiliation{Harvard-Smithsonian Center for Astrophysics, Cambridge,
Massachusetts, 02138, USA} \affiliation{Department of Physics, Harvard
University, Cambridge, Massachusetts, 02138, USA} \affiliation{Lebedev
Institute of Physics, Moscow, 117924, Russia}

\author{R.L. Walsworth}
\affiliation{Harvard-Smithsonian Center for Astrophysics, Cambridge,
Massachusetts, 02138, USA} \affiliation{Department of Physics, Harvard
University, Cambridge, Massachusetts, 02138, USA}

\date{\today}

\begin{abstract}
We demonstrate that first-order light-shifts can be cancelled for an
all-optical, three-photon-absorption resonance
(``\emph{N}-resonance'') on the $D_1$ transition of ${}^{87}$Rb. This
light-shift cancellation enables improved frequency stability for an
\emph{N}-resonance clock. For example, using a table-top apparatus
designed for \emph{N}-resonance spectroscopy, we measured a
short-term fractional frequency stability (Allan deviation) $\simeq
1.5 \times 10^{-11}~\tau^{-1/2}$ for observation times $1~\mathrm{s}
\lesssim \tau \lesssim 50$~s. Further improvements in  frequency
stability should be possible with an apparatus designed as a
dedicated \emph{N}-resonance clock.
\end{abstract}

%
%
%
\maketitle

There is great current interest in developing small, economical
atomic frequency standards (clocks) with fractional frequency
stability $\sim 10^{-12}$ or better. Significant progress toward this
goal has been achieved using coherent population trapping (CPT)
resonances in atomic vapor
\cite{vanier05apb}. However, the frequency stability of CPT clocks is
limited in part
by light-shifts, i.e., shifts of the resonance frequency due to the
applied electromagnetic
fields~\cite{Zhu00,Vanier03}.

Recently our group demonstrated that a three-photon-absorption
resonance (known as an ``\emph{N}-resonance'') is a promising
alternative for small atomic clocks~\cite{zibrov05pra}. Here we show
that it is possible to cancel first-order light-shifts by optimizing
the intensity ratio and frequency of the two optical fields that
create and interrogate the \emph{N}-resonance.  Employing such
light-shift cancellation in a simple, table-top apparatus, we
observed promising short-term frequency stability ($\approx 1.5
\times 10^{-11}~\tau^{-1/2}$) for an \emph{N}-resonance on the $D_1$
transition of ${}^{87}$Rb vapor. We expect superior frequency
stability will be possible in a small \emph{N}-resonance clock
designed for good thermal control, low phase noise, etc.
\begin{figure}[h]
\includegraphics[width=0.8\columnwidth]{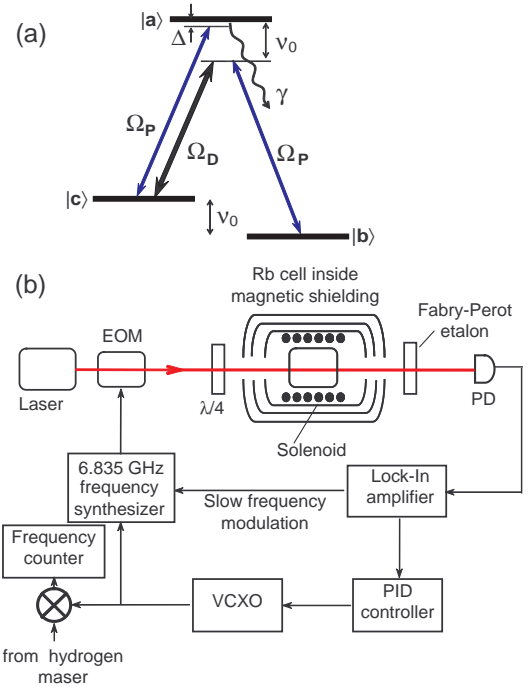}
\caption{ (a) \emph{N}-resonance interaction scheme. $\Omega_P$ and
$\Omega_D$ are the probe and drive optical fields that create and
interrogate the \emph{N}-resonance, $\nu_0$ is the hyperfine
splitting of the two lower energy levels $|b\rangle$ and $|c\rangle$,
$\gamma$ is the collisionally-broadened decoherence rate of the
excited state $|a\rangle$, and $\Delta$ is the one-photon detuning of
the probe field from resonance with the $|c\rangle$ to $|a\rangle$
transition.
(b) Schematic of the experimental setup. See text for abbreviations.}
\label{setup.fig}
\end{figure}

Fig.~\ref{setup.fig}(a) shows the \emph{N}-resonance interaction
scheme~\cite{zibrov05pra,Zibrov-Observation}. A probe field
$\Omega_P$ and drive field $\Omega_D$ are in two-photon Raman
resonance with the ground-state hyperfine levels $|b\rangle$ and
$|c\rangle$, with $\Omega_P$ nearly resonant with the optical
transition $|c\rangle \rightarrow |a\rangle$ and $\Omega_D$
red-detuned from this optical transition by the ground-state
hyperfine splitting $\nu_0$ ($\simeq 6.835$~GHz for ${}^{87}$Rb). The
two-photon Raman process drives atoms coherently from state
$|b\rangle$ to $|c\rangle$, followed by a one-photon transition to
excited state $|a\rangle$ via absorption from field $\Omega_P$.
Together, this three-photon process produces a narrow absorptive
resonance in the probe field transmitted intensity, with a width that
is limited by the relaxation rate of the atoms' ground-state
coherence.

For such an idealized three-level \emph{N}-resonance, the light-shift
$\delta$ (i.e., the detuning from $\nu_0$ of the difference frequency
between the probe and drive fields, as measured by maximum probe
field absorption) consists of three leading (first-order) terms:
shifts of both ground-states due to interaction with the strong,
far-detuned drive field, and a shift of ground-state $|c\rangle$ due
to interaction with the near-resonant probe field:
%
\begin{equation}\label{shift1}
\delta \approx
-\frac{|\Omega_D|^2}{\nu_0+\Delta}+\frac{|\Omega_D|^2}{2\nu_0+\Delta}+\frac{|\Omega_P|^2\Delta}{\Delta^2+\gamma^2/4}.
\end{equation}
Here $\Delta$ is the one-photon detuning of the probe field from
resonance, $\gamma$ is the collisionally-broadened decoherence rate
of the excited state, and $\Omega_P$ and $\Omega_D$ indicate the
probe and drive fields' Rabi frequencies. The light-shifts due to the
far-detuned drive field (the first and second terms in
Eq.(\ref{shift1})) are proportional to the drive field intensity, but
practically independent of the laser frequency for $\Delta \ll
\nu_0$. In contrast, the light-shift due to the near-resonant probe
field (the last term in Eq.(\ref{shift1})) has a strong
dispersive-like dependence on $\Delta$. Thus, near the extrema,
$\Delta = \pm\gamma/2$, the total \emph{N}-resonance light-shift has
only a quadratic dependence on the probe field detuning:
\begin{equation} \label{shift2}
\delta \approx
-\frac{|\Omega_D|^2}{2\nu_0}\pm\frac{|\Omega_P|^2}{\gamma}\mp
\frac{2|\Omega_P|^2}{\gamma^3}(\Delta\mp\gamma/2)^2.
\end{equation}

This light-shift can then be cancelled by (i) detuning the probe
field to the high-frequency extremum, and (ii) properly setting the
intensity ratio of the drive and probe fields:
\begin{equation}\label{ratio}
\Delta=\gamma/2 , \ \ \frac{|\Omega_P|^2}{|\Omega_D|^2} =
\frac{\gamma}{2\nu_0}.
\end{equation}
With such light-shift cancellation, the measured \emph{N}-resonance
center frequency should be insensitive (to leading order) to
fluctuations of the probe field frequency and total laser intensity.
Note that the light-shift cancellation does not depend on the
absolute values of either optical field.
\begin{figure}
\includegraphics[width=0.9\columnwidth]{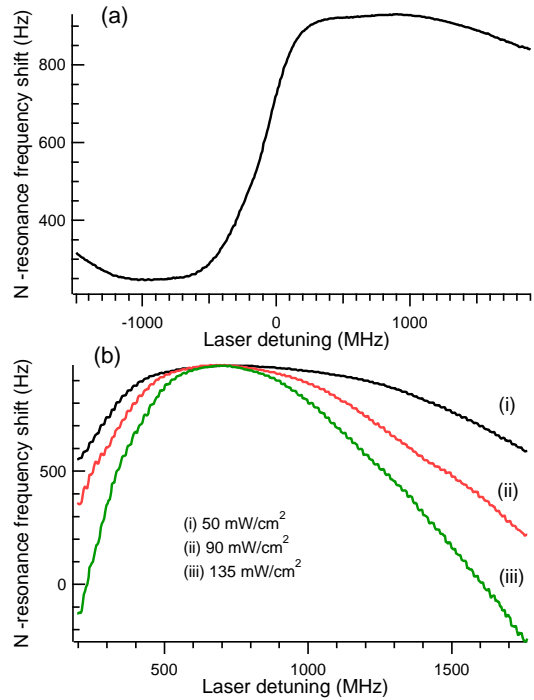}
\caption{Measured detuning light-shift, i.e., \emph{N}-resonance
frequency shift as a function of the detuning $\Delta$ of the probe
field from the $5S_{1/2}~F=2 \rightarrow 5P_{1/2}~F'=2$ transition in
${}^{87}$Rb vapor. (a) Example of the light-shift's dispersive-like
dependence on $\Delta$; total laser intensity $\simeq$ 30 mW/cm$^2$.
(b)  Insensitivity of the light-shift to variations in the probe
field frequency and total laser intensity near the optimized probe
field detuning ($\approx 700$~MHz); probe/drive intensity ratio
$\simeq 11\%$, to cancel intensity light-shift (see
Fig.~\ref{shiftint.fig}). Buffer gas collisions shift the
\emph{N}-resonance frequency by $\approx$ 500 Hz for the data shown
here.} \label{shiftdet.fig}
\end{figure}
%
\begin{figure}[h]
\includegraphics[width=0.9\columnwidth]{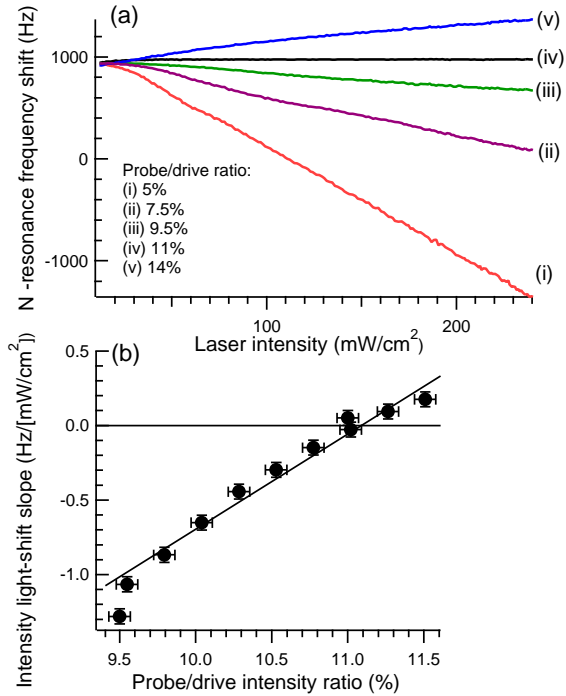}
\caption{(a) Measured intensity light-shift, i.e., \emph{N}-resonance
frequency shift as a function of laser intensity for different ratios
between the probe and drive field intensities; probe field  detuned
to the light-shift maximum ($\Delta \approx 700$~MHz).
(b) Fitted linear slopes for the measured light-shift variation with
laser intensity, for probe/drive field intensity ratios near the
intensity light-shift cancellation value.} \label{shiftint.fig}
\end{figure}

To verify these predictions, we measured ${}^{87}$Rb
\emph{N}-resonance light-shifts  using the experimental setup shown
in Fig.~\ref{setup.fig}(b). We phase-modulated the output of a
free-running New Focus external cavity diode laser using an
electro-optical modulator (EOM), which produced two optical sidebands
separated by $\simeq 6.835$~GHz. The EOM was driven by a microwave
synthesizer locked to a 100 MHz voltage-controlled crystal oscillator
(VCXO). The laser frequency was adjusted such that the high-frequency
sideband (serving as the probe field $\Omega_P$) was tuned close to
the $5S_{1/2}~F=2 \rightarrow 5P_{1/2}~F'=2$ transition of
${}^{87}$Rb ($\lambda\simeq 795$~nm); the carrier-frequency field
then served as the drive field $\Omega_D$. The probe/drive field
intensity ratio was set by the EOM phase-modulation index. The laser
beam was circularly polarized using a quarter wave plate and weakly
focused to a diameter of 0.8~mm before entering the Rb vapor cell.

We employed a cylindrical Pyrex cell containing isotopically enriched
${}^{87}$Rb and a mixture of buffer gases (15~Torr Ne + 15~Torr Ar +
5~Torr \A) chosen to minimize the temperature dependence of the
${}^{87}$Rb ground-state hyperfine frequency shift due to buffer gas
collisions~\cite{Vanier-book}. Associated collisional broadening of
the excited state is estimated to be $\gamma\approx \pi\times
1.2$~GHz. During experiments, the vapor cell was heated to
55\,${}^\circ$C and isolated from external magnetic fields with three
layers of high permeability shielding. A small ($\approx 10$~mG)
longitudinal magnetic field was applied to lift the degeneracy of the
Zeeman sublevels and separate the desired $F=1$, $m_F=0$ to $F=2$,
$m_F=0$ clock transition (no first-order magnetic field dependence)
from the $m_F=\pm1$ transitions (first-order Zeeman splitting).
The strong drive field and the lower-frequency sideband were filtered
from the light
transmitted through the cell using a quartz, narrow-band Fabry-Perot etalon
(free spectral range of 20~GHz, finesse of 30), which was tuned to
the frequency of the
probe field and placed before the photodetector (PD).

To lock the frequency of the VCXO (and hence the detuning of the
probe and drive fields) to the \emph{N}-resonance, we superimposed a
slow frequency modulation at $f_m=400$~Hz on the 6.8~GHz signal from
the microwave synthesizer. We demodulated the photodetector output at
$f_m$ with a lock-in amplifier, and used the in-phase lock-in
amplifier output as an error signal to feed back to the VCXO. We then
monitored the frequency of the locked VCXO (and thus the
\emph{N}-resonance center frequency) by comparing it with a $100$~MHz
signal derived from a hydrogen maser.

%
Figs.~\ref{shiftdet.fig} and \ref{shiftint.fig} show examples of the
measured dependence of the \emph{N}-resonance frequency on laser
detuning, intensity, and probe/drive field intensity ratio.
Consistent with Eq.(\ref{shift1}), we observed two extrema in the
detuning light-shift: one below and one above the probe field
resonance frequency; see Fig.~\ref{shiftdet.fig}(a). As illustrated
in Fig.~\ref{shiftdet.fig}(b), we found that the probe field detuning
at the light-shift maximum ($\Delta \approx 700$~MHz) is effectively
independent of the total laser intensity, as expected from
Eq.(\ref{shift2}), when the probe/drive field intensity ratio
$|\Omega_P|^2/|\Omega_D|^2$ is set to make the total light-shift
independent of the laser intensity. We determined this intensity
light-shift cancellation ratio ($|\Omega_P|^2/|\Omega_D|^2 \simeq
11\%$, given by EOM phase-modulation index $\simeq 0.22$) from the
measurements shown in Fig.~\ref{shiftint.fig}. Note that the
experimentally-determined ratio is in reasonable agreement with the
prediction of $|\Omega_P|^2/|\Omega_D|^2 \simeq 9\%$ given by
Eq.(\ref{ratio}) for our experimental conditions.

%
\begin{figure}
\includegraphics[width=0.9\columnwidth]{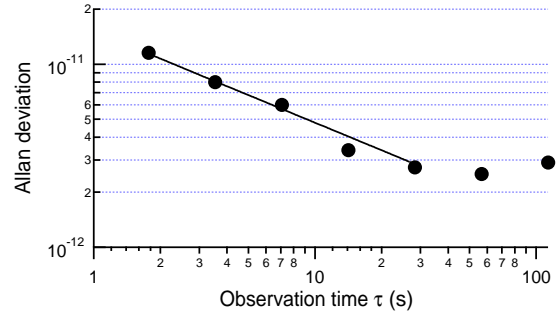}
\caption{Measured frequency stability of a 100 MHz crystal oscillator
locked to the ${}^{87}$Rb \emph{N}-resonance, relative to a hydrogen
maser.
} \label{allanvar.fig}
\end{figure}

We next characterized the frequency stability of a crude
``\emph{N}-resonance clock'' --- i.e., the VCXO locked to the
${}^{87}$Rb \emph{N}-resonance as described above --- relative to a
hydrogen maser. For this measurement we tuned our system to the
conditions for optimal light-shift cancellation (laser detuning
$\Delta \simeq 700$~MHz, probe/drive intensity ratio $\simeq 11\%$)
with total laser power $\simeq 140~\mu$W (intensity $\simeq$ 30
mW/cm$^2$). Under such conditions the \emph{N}-resonance linewidth
$\simeq 1400$~Hz (FWHM) and contrast $\simeq 7\%$, which implies a
shot-noise-limited short-term frequency $\simeq 5\times
10^{-14}~\tau^{-1/2}$~\cite{Vanier03}. Fig.~\ref{allanvar.fig} shows
the measured \emph{N}-resonance clock fractional frequency stability
(Allan deviation). The short-term stability $\simeq 1.5 \times
10^{-11}~\tau^{-1/2}$ for observation times $1~\mathrm{s} \lesssim
\tau \lesssim 50$~s. At longer times the stability degrades due to
uncontrolled temperature and mechanical variations in our table-top
apparatus, as well as long-term drifts of the laser frequency.
Despite this non-optimal clock apparatus, the short-term
\emph{N}-resonance frequency stability is already better than that
provided by many recently-demonstrated CPT
clocks~\cite{vanier05apb,Finland,knappe05oe}. We expect that both the
short- and long-term \emph{N}-resonance frequency stability can be
further improved by straightforward optimization of the VCXO
lock-loop (to reduce phase noise), temperature stabilization, laser
control, etc. We also expect that a high-stability \emph{N}-resonance
clock should be possible in a compact physical package (with vapor
cell volume $\sim$ 1 mm$^3$), because of promising \emph{N}-resonance
characteristics at high buffer gas pressure~\cite{zibrov05pra}.

In conclusion, we demonstrated cancellation of first-order
light-shifts for an all-optical, three-photon-absorption
\emph{N}-resonance on the $D_1$ line of ${}^{87}$Rb vapor. Employing
this light-shift cancellation in a table-top apparatus not engineered
for stable clock performance, we nonetheless observed
\emph{N}-resonance frequency stability comparable to or better than
existing CPT clocks. Signficant improvements in \emph{N}-resonance
frequency stability should be possible in a small device with
standard techniques. We note also that similar light-shift
cancellation is possible for other \emph{N}-resonances, e.g., the Rb
$D_2$ line ($\lambda=780$~nm). Currently, diode lasers for the $D_2$
line of Rb and Cs are more easily obtained than for the $D_1$ line.

The authors are grateful to J.\ Vanier, M.\ D.\ Lukin, and V.\ L.\ Velichansky
for useful discussions.  This work was supported by ONR, DARPA and the
Smithsonian Institution. A.\ V.\ T.\ and V.\ I.\ Y.\ acknowledge support from
RFBR (grants no. 05-02-17086 and 04-02-16488).

\end{document}